# Finding Local Experts for Dynamic Recommendations Using Lazy Random Walk

Diyah Puspitaningrum
Dept. of Computer Science
University of Bengkulu
Bengkulu, Indonesia
diyahpuspitaningrum@gmail.com

Julio Fernando
Dept. of Computer Science
University of Bengkulu
Bengkulu, Indonesia
juliofernando7@gmail.com

Edo Afriando
Dept. of Computer Science
University of Bengkulu
Bengkulu, Indonesia
afriando.edo94@gmail.com

Ferzha Putra Utama
Dept. of Computer Science
University of Bengkulu
Bengkulu, Indonesia

Rina Rahmadini
Dept. of Computer Science
University of Bengkulu
Bengkulu, Indonesia

Y. Pinata
Dept. of Computer Science
University of Bengkulu
Bengkulu, Indonesia

*Abstract*—Statistics based privacy-aware recommender systems make suggestions more powerful by extracting knowledge from the log of social contacts interactions, but unfortunately, they are static — moreover, advice from local experts effective in finding specific business categories in a particular area. We propose a dynamic recommender algorithm based on a lazy random walk that recommends top-rank shopping places to potentially interested visitors. We consider local authority and topical authority. The algorithm tested on FourSquare shopping data sets of 5 cities in Indonesia with k-steps={5,7,9} (lazy) random walks and compared the results with other state-of-the-art ranking techniques. The results show that it can reach high score precisions (0.5, 0.37, and 0.26 respectively on p@1, p@3, and p@5 for k=5). The algorithm also shows scalability concerning execution time. The advantage of dynamicity is the database used to power the recommender system; no need to be very frequently updated to produce a good recommendation.

## I. Introduction

There have been studies that show that people are likely to prefer learning from local experts who know the neighborhood well and have first-hand experience [1]. In the case of shopping place recommender systems, to predict a user u's preference for place p, the recommender model should be built from an appropriate local expert given a database of local shopping places with positive reviews. Using statistics based model has a drawback that it leads to choosing the same most popular spot by visitors reviews. In the case of our proposed lazy random walk recommender model, a recommender system recommend a place because the user listens either to recommendations given by his social contacts about a specific region or to suggestions given by local experts from the global network, with both of them must meet the local authority and topical authority criteria.

Furthermore, finding local experts is not a trivial task. Since FourSquare data only comes from visitors' reviews of some shopping places, the data is sparse for both users and shopping places. Typical shopping place on FourSquare has few visitors' reviews.

This work aims to come up with a system that can find dynamically local experts, to produce dynamic recommendations, given a query about shopping in certain cities in which suggestions from a local expert from a friend's graph is more likely than the one from global networks.

The rest of the paper organized as follows: Section 2 describes the problem definition. All related works described in Section 3. The details of the proposed model in section 4. Section 5 presented the experimental results and comparison with existing methods. Finally, we conclude the paper in section 6 and introduce some directions for future research.

## II. Problem Definition

In privacy-aware recommender systems we have a set of users $U = \{u_1, ..., u_N\}$ assuming that $N$ is the total number of users in the shopping dataset and a set of shopping places $P = \{p_1, ..., p_M\}$ and $M$ is the total number of shopping places by local expert's query category $c(q)$. Each user $u_i$ can be associated with a location $l_i$ and a set of categories $C_i$. The objective is to find the collection of shopping places from users $U(q)$ such that for each $u_j$ in $U(q)$, $c(q)$ is in $C_j$, and $l(q)$ is local to $l_j$, dynamically using Lazy Random Walks. By privacy-aware, we prefer recommendations from social contacts rather than from global users.

Several main definitions as follows:

1. *Local authority*: A user is said a candidate of a local expert if the user's location, and the query location, are in the same city.

2. *Topic authority*: A user is said a candidate of a local expert if the user has considerable knowledge about the query category to be recognized as an expert or not. Since the dataset is sparse, a user is a local expert candidate if he has reviews about category $c(q)$.

3. *Places*: Recommendations are generated from shopping places with the majority of positive reviews and suggested by a local expert that meets both the *local authority* and the *topical authority* criteria. In this work, we do sentiment analysis first to categorize shopping places dataset into two bins: positive and negative bins.

## III. Related Works

In this section, we review related work on recommender systems related to the use of local experts and random walk. Tanvi Jindal (2015) [2] solves the problem of finding local experts from the Yelp dataset. He proves that the random forest algorithm works best to classify expert and non-expert, so the recommender system gives the best prediction accuracy. Once we have the set of local experts for a category in a location, we can use it to generate a summary for the businesses in that category. Since they are knowledgeable and influential, using just their reviews would give a higher quality summary about a business.

Torres et al. (2012) [3] use random walks as a query suggestion method to help children find keywords that are more likely to be relevant for them. Abbasssi and Mirrokni (2007) [16] apply the local partitioning algorithms based on refined random walks to approximate the personalized PageRank vector for the use of a recommender system for weblogs based on the link structure among them. In (Bogers, 2010) [14], random walks over the contextual graph models the browsing process of a user on a movie database website. Gori and Pucci (2006) [12] use a random-walk based scoring algorithm to rank products according to expected user preferences. In (Jamali and Ester, 2009), [15] a random walk algorithm is proposed to recommend items in a trusted network. The algorithm recommends items based on ratings expressed by trusted friends, using random walk and probabilistic item selection.

IV. RECOMMENDER MODEL

The strategy of our Lazy Random Walk recommender system when implemented to help for finding local experts as follows: if a user is a person without detected social contacts then use a global graph search for finding local expert, otherwise use a graph derived from his social connections to do the local expert finding.

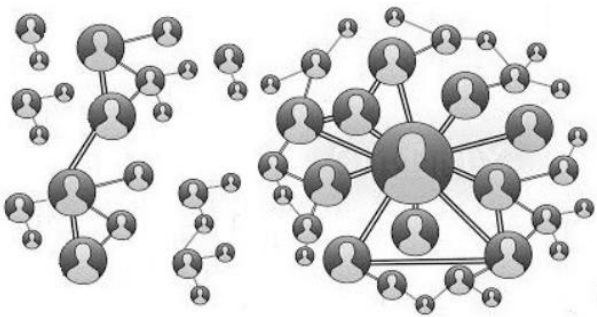

Fig. 1. An illustration of the global view of user's data in the shopping dataset.

Our proposed recommendation system as follows:

1. Query identification: query $q$ (or *category* such as: "shopping mall", "department store", "supermarket", "bookstore", "market"), user id, city, and algorithm.

2. Filter the visitor review data only to a graph that meets the query criteria. Select only reviews that originated from user id that located in the city stated in point 1.

3. If a user is a person without social contacts, then use a global graph search local expert. Otherwise, use a graph derived from his social contacts to do the local expert finding.

4. As we use network properties and random walk variants to test our system, then if method="PageRank" (or other network centrality measures, e.g., "betweenness," "closeness," and "degree"), for the highest local expert score, find his recommendation of places that meet the query. Else: If method="Lazy Random Walk" or "Random Walk," for source node of all users in the graph (see point 3) with walking-steps = {5, 7, 9} use the last step of the walk as a local expert candidate and then find his recommendation of places that meet $q$.

5. Sort recommendations by the number of positive reviews in its descending order.

Remark 1. The "Global" Graph Search Procedure

1. Build a global graph(s) of reviewers (users who have comments on a shopping place that is meet the user query q) by finding each reviewer's contact until two intermediate nodes. A reviewer in this graph can be connected or isolated, so it is possible to have more than one global graph.

2. Since a local expert should influence his local network (in his city), find the highest degree graph as the origin of candidates of local experts.

3. Do lazy random walk or PageRank on the graph with the number of walking steps = $\{5,7,9\}$.

4. The last visited node is the local expert.

Remark 2. The "Privacy-Aware" Graph Search Procedure

1. Build a global graph until two intermediate nodes ahead of a user who request the query.

2. Do lazy random walk on the graph.

3. The last visited node is the local expert.

V. EXPERIMENTS

A. *Experimental Design*

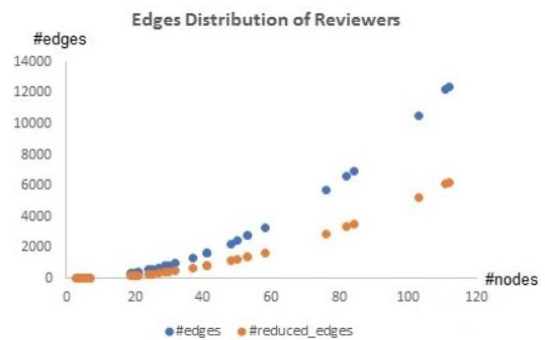

2(a). Edge Distribution of User Review Counts

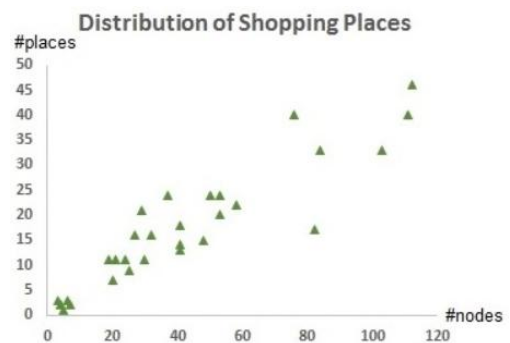

2(b). Places Distribution of Local Expert Candidates

Fig. 2. Statistics of dataset

We worked with the FourSquare shopping places data set from 5 cities in Indonesia. It consists of 1) 176 shopping places data from 5 cities in Indonesia. 2) 3844 visitors' reviews trained using Support Vector Machine or ensemble backpropagation to classify positive, neutral, as well as the negative class of shopping places reviews (sentiment analysis tasks). 3) 14309 users data.

Fig. 2 shows the statistics of the dataset. Fig. 2(a) illustrates the edge distribution of the number of reviewers (#nodes) compare to edges from the global graph (#edges) and edges from the highest degree graph (#reduced edges). Fig. 2(b) shows the distribution of shopping places compare to the number of reviewers.

We have 220 queries $q_i$ consists of combinations of: query+city+mode+$k$-walking steps. query = {"mall", "inexpensive market", "discount books", "dress shop", "comfortable shopping", "complete and inexpensive shopping"}. city loc= {Jakarta, Bandung, Yogyakarta, Palembang, Bengkulu}. mode = {"global", "pa"}. $k$-walking steps = {5,7,9,11,13,15,17,19,21,23} walking steps. The term "*global*" refers to *global* graph search procedure; the "*pa*" stands for *privacy aware* graph search procedure. Due to we have less data for privacy aware networks, the *pa* mode is replaced by the global networks.

To be able to generate a recommendation, the recommender system must be trained to classify positive and negative reviews of shopping places dataset correctly. We test our system to do this sentiment analysis task using five different classifiers: backpropagation neural nets that run individually (bp_1, bp_2, bp_3), and an ensemble backpropagation which runs bp_1, bp_2, bp_3 simultaneously in ensemble mode, and an SVM classifier. We use ensemble since combining multiple classifiers leads to better performance than using only an individual one.

The bp_1 has learning rate=0.00005, 100 hidden layers, maximum iterations of 100 iterations. The bp_2 has learning rate=0.0001, 200 hidden layers, maximum iterations=500. The bp_3 has a learning rate=0.00015 and 300 hidden layers with maximum iterations of 1000 iterations. We use the SVM classifier with nonlinear degree 3 of RBF kernel and C-support vector classification parameter equal to 3 with a maximum 1000 iteration. A large C gives the system low bias and high variance.

We implemented our algorithm and other methods in Python. We used an Intel Core i7 2.6GHz CPU with 8GB RAM to run our experiments on a Windows 10 system.

Sulieman et al. (2013) [9] proposes an algorithm on a semantic relevance measure and social network centrality measures for the recommendation. Therefore for evaluation, we compare our Lazy Random Walks (labels to *lrw*) to the other five methods: Random Walks (*rw*), PageRanks (*pr*), and network centralities measurements such as betweenness (*btw*), closeness (*cls*), and degree (*deg*).

*B. Evaluation Metrics*

In our experiments, we compare the results for different methods. Since we can view generating recommendations as a kind of link prediction task, we use *R*-score (Breese et al., 1998) [18] and average over links to handle the extreme case of scarce local experts and top-rated local experts (Singh et al., 2007)[10]. By a user's likelihood: $p(u,t) = 2^{-\frac{(j_t-1)}{(\lambda-1)}}$ the $j_t$ is the position of a shopping place $t$ on a ranked list of statistic-based shopping places, which has the most positive reviews in decrease order. The $\lambda$ is equal to the number of walking steps (or walking steps for short) divide by 2 (a position in the list with 50-50 stopping chance). As one of the evaluation tools, we compare two parameters of *R*-score: the *R*] that each, respectively, is an expected utility and the maximum possible R-score.

Given recommendation places from a local expert, *R*-score computes:

$$R_u = \sum_{i=1}^{n} p(u,g_i) \bigcup (u_i, g_i) \quad (1)$$

Where *n* = number of places with $R_u^{max}$ is a maximum $p(u,g_i)\bigcup(u_i,g_i)$. A utility function to predict whether a local expert candidate has a non-popular shopping place defined as

$$U(u_i, g_i) = -\log\left(\frac{1}{len(nodes)} * walking\_steps\right) \quad (2)$$

where *len*(*nodes*) is the number of local expert candidates.

For validity, the maximum possible *R*-score is achieved by

$$E[R_u] = \frac{1}{m}\sum \frac{R_u}{R_u^{max}} \quad (3)$$

In our experiment, we use *m*=10 as we generate ten times of recommendation for each user query.

We use $R_u$ and $E[R_u]$ to compute the mean squared error (MSE). Here we defined MSE as the average of the squares of the discrepancies between the expected utility ($R_u$) and the maximum achieved possible *R*-score ($E[R_u]$) of Random Walk variant. An excellent recommendation system must have low MSE.

We also compute precision at *x*, or *p@x*, as the second metric where *x*={1,3,5,10,15,20,30} with value range of [0; 1]. *p@x* defined as the number of correct recommendations over x gold standard recommendation. A gold standard dataset generated by comments or user reviews about shopping places that meet user query criteria and sort in decrease order based on the most positive reviews.

Further, to measure scalability, we compute the computational cost of time for building graph ($t_{graph}$), time for running random walker ($t_{algo}$), time for others required for producing output from recommender systems ($t_{other}$), and total execution time ($t_{total} = t_{graph} + t_{algo} + t_{other}$). An online system needs a slightly fast running time.

*C. Experimental Results*

TABLE I shows the results of the sentiment analysis task. SVM shows the best accuracy both in seen instances (system validation) and unseen instances (system testing).

TABLE I. SENTIMENT ANALYSIS ON USER REVIEWS IN SHOPPING PLACES DATASET

| Classifier | System Validation (%) | System Testing (%) |
|---|---|---|
| bp_1 | 100 | 96.45 |
| bp_2 | 100 | 96.45 |
| bp_3 | 100 | 96.58 |
| ensemble_bp | 100 | 96.45 |
| Svm kernel rbf degree 3 | 100 | 100 |

TABLE II shows precisions for all methods used in this paper. Both Random Walk and Lazy Random Walk show useful precisions compare to recommendations using PageRank and other network properties. From Fig. 3 the Lazy Random Walk with walking steps (or $k$-steps)=5, in general, outperform other methods followed by the Random Walk method with also has five walking steps.

TABLE II.  PRECISIONS @$x$ OF SHOPPING PLACES DATASET

| p@1 | 1 | 2 | 3 | 4 | 5 | 6 |
|---|---|---|---|---|---|---|
| betweenness | 0.4 | 0.6 | 0.4 | 0.4 | 0.2 | 0.2 |
| closeness | 0.4 | 0.6 | 0.4 | 0.4 | 0.2 | 0.2 |
| degree | 0.4 | 0.6 | 0.4 | 0.4 | 0.2 | 0.2 |
| pagerank | 0.2 | 0.6 | 0.2 | 0.2 | 0.2 | 0.2 |
| rw k=5 | 0.34 | 0.36 | 0.46 | 0.38 | 0.28 | 0.38 |
| rw k=7 | 0.13 | 0.15 | 0.15 | 0.23 | 0.08 | 0.15 |
| rw k=9 | 0.08 | 0.25 | 0.15 | 0.25 | 0.16 | 0.1 |
| lrw k=5 | 0.5 | 0.32 | 0.5 | 0.44 | 0.3 | 0.26 |
| lrw k=7 | 0.35 | 0.2 | 0.13 | 0.53 | 0.1 | 0.13 |
| lrw k=9 | 0.35 | 0.25 | 0.1 | 0.55 | 0.1 | 0.06 |
| **p@3** | *1* | *2* | *3* | *4* | *5* | *6* |
| betweenness | *0.25* | *0.25* | *0.5* | *0.33* | *0.2* | *0* |
| closeness | *0.25* | *0.33* | *0.5* | *0.33* | *0.2* | *0* |
| degree | *0.25* | *0.33* | *0.5* | *0.33* | *0.2* | *0* |
| pagerank | *0.17* | *0.25* | *0.33* | *0.33* | *0.27* | *0* |
| rw k=5 | *0.18* | *0.26* | *0.26* | *0.36* | *0.24* | *0.19* |
| rw k=7 | *0.16* | *0.26* | *0.29* | *0.18* | *0.09* | *0.19* |
| rw k=9 | *0.18* | *0.28* | *0.26* | *0.2* | *0.13* | *0.17* |
| lrw k=5 | *0.17* | *0.27* | *0.27* | *0.37* | *0.21* | *0.12* |
| lrw k=7 | *0.19* | *0.28* | *0.24* | *0.24* | *0.08* | *0.12* |
| lrw k=9 | *0.2* | *0.27* | *0.23* | *0.28* | *0.08* | *0.12* |
| **p@5** | *1* | *2* | *3* | *4* | *5* | *6* |
| betweenness | *0.25* | *0.25* | *0.5* | *0.1* | *0.05* | *0.15* |
| closeness | *0.25* | *0.25* | *0.5* | *0.1* | *0.05* | *0.15* |
| degree | *0.25* | *0.25* | *0.45* | *0.1* | *0.05* | *0.15* |
| pagerank | *0.1* | *0.2* | *0.5* | *0.15* | *0.05* | *0.15* |
| rw k=5 | *0.2* | *0.25* | *0.25* | *0.23* | *0.11* | *0.19* |
| rw k=7 | *0.2* | *0.25* | *0.25* | *0.19* | *0.1* | *0.21* |
| rw k=9 | *0.2* | *0.27* | *0.25* | *0.18* | *0.1* | *0.15* |
| lrw k=5 | *0.21* | *0.2* | *0.26* | *0.22* | *0.07* | *0.16* |
| lrw k=7 | *0.21* | *0.26* | *0.26* | *0.21* | *0.07* | *0.14* |
| lrw k=9 | *0.21* | *0.21* | *0.25* | *0.22* | *0.06* | *0.15* |

a. Queries: 1: dress shop ; 2: discount books; 3: mall; 4: inexpensive market; 5: complete and inexpensive shopping; 6: comfortable shopping.
b. Methods: rw: random walk; lrw: lazy random walk.

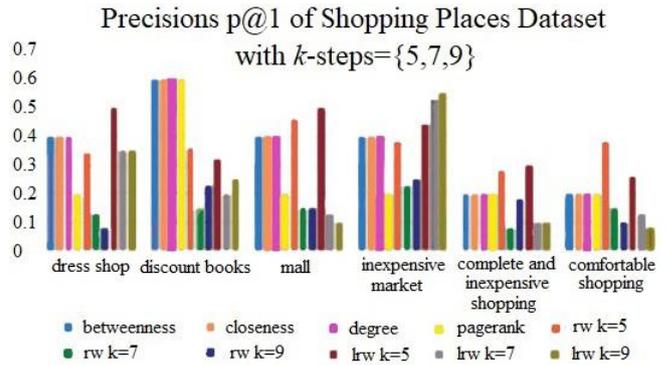

Fig. 3. P@1 of shopping places dataset

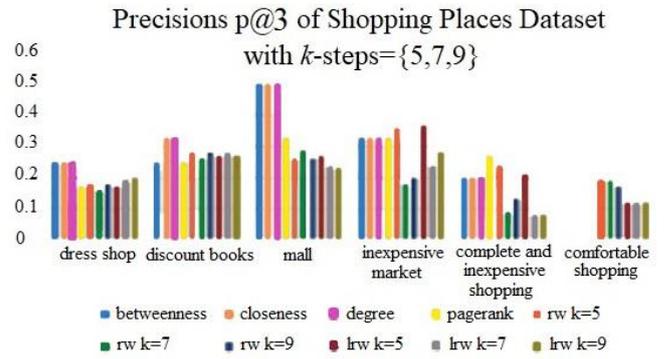

Fig. 4. P@3 of shopping places dataset

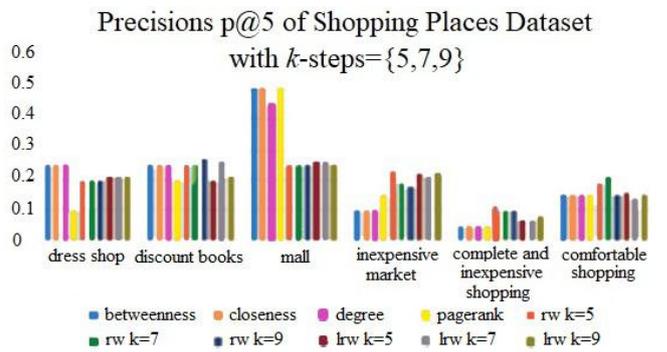

Fig. 5. P@5 of shopping places dataset

However, the precision of Random Walk and Lazy Random Walk decrease on p@3 and p@5, as shown in Fig. 4 and Fig. 5, where the network properties method (including the PageRank) indicates better performance than the Random Walk variants. The rationale behind it is because of the random character of Random Walk and Lazy Random Walk, whereas, in network properties, we use a static chosen highest score node of network properties as the local expert. Therefore recommendation systems using network properties method always produce the same recommendation results for the same user query.

TABLE III shows an example of recommendation results given user query = "mall" and city = "Bandung." The figure shows that the results are reliable. The gold standards are the ensemble backpropagation for the Random Walk technique and the SVM classifier for the Lazy Random Walk technique because the experiments set up was running separately for the two methods. The different gold standards are as results of the different classifiers used in sentiment analysis.

TABLE III. A QUERY EXAMPLE FROM FOURSQUARE LOG QUERIES. QUERY="MALL", CITY="BANDUNG", MODE="GLOBAL", ORDER BY HIGHEST POSITIVE COMMENTS AND LOWEST NEGATIVE COMMENTS

| Query | Mall, city="Bandung", mode="global" |
|---|---|
| lrw + SVM | [('Istana Plaza (IP)',24L,5L), ('LotteMart',23L,3L), ('Gramedia',20L,3L)] |
| rw + SVM | [('Istana Plaza (IP)',24L,5L), ('LotteMart',23L,3L), ('METRO Department Store',21L,6L)] |
| SVM gold standard | [('Istana Bandung Electronic Center (BEC)',26L,1L), ('Trans Studio Mall (TSM)',24L,4L), ('Istana Plaza (IP)',24L,5L), ('Cihampelas Walk (CiWalk)',23L,1L), ('Lotte Mart',23L,3L), ('Braga City Walk',23L,4L), ('Bandung Trade Centre - BTC Fashion Mall',23L,3L), ('Bandung Indah Plaza (BIP)',21L,2L), ('riaujunction',21L,5L), ('METRO Department Store',21L,6L), ('Gramedia',20L,3L), ('Festival Citylink',19L,5L), ('Paris Van Java (PVJ)',18L,3L), ('Liana swalayan',5L,0L), ('Living Plaza Dago',3L,0L)] |
| btw + ens. bp | [('Trans Studio Mall (TSM)',27L,1L), ('Istana Plaza (IP)',25L,4L), ('Festival Citylink',21L,3L)] |
| cls + ens. bp | [('Trans Studio Mall (TSM)',27L,1L), ('Istana Plaza (IP)',25L,4L), ('Festival Citylink',21L,3L)] |
| deg + ens. bp | [('Trans Studio Mall (TSM)',27L,1L), ('Istana Plaza (IP)',25L,4L), ('Festival Citylink',21L,3L)] |
| pr + ens. bp | [('Trans Studio Mall (TSM)',27L,1L), ('Istana Plaza (IP)',25L,4L), ('Festival Citylink',21L,3L)] |
| ens. bp gold standard | [('Setiabudhi Supermarket',29L,1L), ('Trans Studio Mall (TSM)',27L,1L), ('Istana Bandung Electronic Center (BEC)',26L,1L), ('Istana Plaza (IP)',25L,4L), ('Bandung Trade Centre - BTC Fashion Mall',25L,1L), ('Toko Buku Togamas',24L,4L), ('riaujunction',24L,2L), ('METRO Department Store',24L,3L), ('Gramedia',22L,1L), ('Braga City Walk',22L,5L), ('Cihampelas Walk (CiWalk)',21L,3L), ('Festival Citylink',21L,3L), ('Bandung Indah Plaza (BIP)',21L,2L), ('Paris Van Java (PVJ)',20L,1L), ('LotteMart',19L,7L), ('Liana swalayan',5L,0L)] |

*lrw*: lazy random walk; *rw*: random walk; *SVM*: Support Vector Machines; *bp*: backpropagation; *ens*: ensemble; *btw*: betweenness, *deg*: degree, *pr*: pagerank; *cls*: closeness graph.

('LotteMart',23L,3L) means the query result has twenty three positive comments and three negative comments.

We investigate further the performance of Random Walk and Lazy Random Walk systems during the walking process through *k*-steps (*k* = {5,7,9,11,13,15,17,19,21,23}). From Fig. 6, we can see that, in general, the performance of the system decreases through the steps. The performance of the methods is excellent only under small steps (in Fig. 6 on *k*-steps={5,7,9} with MSE-scores$\leq$0.206). The more levels we have can lead us to an unpopular local expert, which further can yield to the unpopular places.

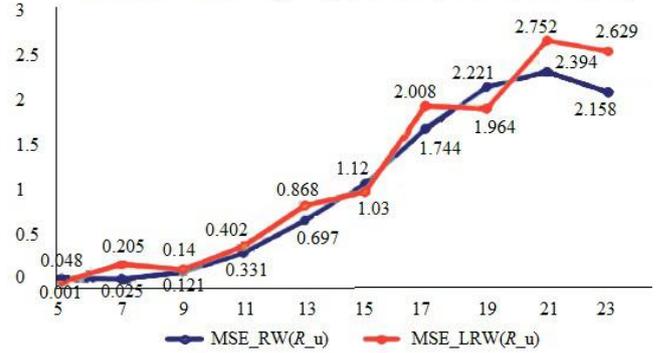

Fig. 6. MSE

About systems scalability, for both of Lazy Random Walk (LRW) and Random Walk (RW), most of the time (93%) was consumed in t(graph stage), while the 3% of the time is on talgo graph (the RW/LRW stage). The remaining time (4%) was for t (output or graphical user interface systems). Our recommendation systems itself is very scalable that it needs only a total of 3.215 seconds for the random walk and 3.222 seconds for the lazy random walk. The time difference between the two methods is insignificant due to the small size of our used dataset.

From the results in TABLE II, the precisions of Lazy Random Walk and the Random Walk can reach sufficiently high scores (0.5, 0.37, and 0.26 for *p@1*, *p@3*, and *p@5* for Lazy Random Walk and 0.46, 0.36, and 0.25 for *p@1*, *p@3*, and *p@5* for Random Walk). Moreover, since they are also slightly fast, then both methods are potential for online implementation, embedded in a recommendation system. Another advantage, because it is dynamic, any recommendation systems that utilized either Random Walk or Lazy Random Walk method are natural to be maintained due to their databases no need to be very frequently updated to produce useful recommendations.

## VI. CONCLUSION

Both Lazy Random Walk and Random Walk methods are potential for online implementation in a recommendation system since they offer slightly fast running time with reliable results. From the industry perspective, both techniques provide easiness viz. infrequent updating time but still produce a good recommendation.

ACKNOWLEDGMENT

Thanks to Hanifah and Y. Setiawan for their comments. Also, thanks to students: Julio Fernando and Edo Afriando for the video; Rina Rahmadini, Y. Pinata, and P.A. Wicaksono for the datasets.